\documentclass[aps,prb,twocolumn,floats]{revtex4}
\usepackage{bm}
\usepackage{graphicx}

\begin{document}
\title{Tuning valley polarization in a WSe$_2$ monolayer with a tiny magnetic field}

\author{T.~\surname{Smole\'nski}$^{1}$}
\author{M.~\surname{Goryca}$^{1}$}
\author{M.~\surname{Koperski}$^{1,2}$}
\author{C.~\surname{Faugeras}$^{2}$}
\author{T.~\surname{Kazimierczuk}$^{1}$}
\author{K.~\surname{Nogajewski}$^{2}$}
\author{P.~\surname{Kossacki}$^{1}$}\email{Piotr.Kossacki@fuw.edu.pl}
\author{M.~\surname{Potemski}$^{2}$,}\email{Marek.Potemski@lncmi.cnrs.fr}

\affiliation{
$^{1}$ Institute of Experimental Physics, Faculty of Physics, University of Warsaw, ul. Pasteura 5, 02-093 Warsaw, Poland\\
$^{2}$ Laboratoire National des Champs Magn\'etiques Intenses, CNRS-UGA-UPS-INSA-EMFL, 25 Rue des Martyrs, Grenoble 38042, France}
\maketitle

\noindent
{\textbf{In monolayers of semiconducting transition metal dichalcogenides, the light helicity ($\sigma^+$ or $\sigma^-$) is locked to the valley degree of freedom\cite{Xiao_PRL_2012, Cao_NatCommun_2012, Zeng_NatNano_2012, Mak_NatNano_2012, Jones_NatNano_2013}, leading to the possibility of optical initialization of distinct valley populations. However, an extremely rapid valley pseudospin relaxation (at the time scale of picoseconds) occurring for optically bright (electric-dipole active) excitons\cite{Qinsheng_ACSNano_2013, Cong_NanoLett_2014, Wang_PRB_2014, Zhu_PRB_2014} imposes some limitations on the development of opto-valleytronics. Here we show that inter-valley scattering of excitons can be significantly suppressed in a WSe$_2$ monolayer, a direct-gap two-dimensional semiconductor with the exciton ground state being optically dark. We demonstrate that the already inefficient relaxation of the exciton pseudospin in such system can be suppressed even further by the application of a tiny magnetic field of $\sim$100~mT. Time-resolved spectroscopy reveals the pseudospin dynamics to be a two-step relaxation process. An initial decay of the pseudospin occurs at the level of dark excitons on a time scale of 100~ps, which is tunable with a magnetic field. This decay is followed by even longer decay ($>1$~ns), once the dark excitons form more complex objects allowing for their radiative recombination. Our finding of slow valley pseudospin relaxation easily manipulated by the magnetic field open new prospects for engineering the dynamics of the valley pseudospin in transition metal dichalcogenides.}}

Interband excitation of a semiconductor brings up the possibility to transfer the angular momentum of circularly polarized photons to photoexcited carriers, to create a non-equilibrium orientation of their spins and, eventually, to examine the conservation of this orientation in the crystal by probing the polarization degree of the emitted light. This so-called optical orientation has been widely explored in zinc-blende semiconductors with respect to the angular momentum of electronic spins\cite{optical_orientation_1984,spin_physics_2008}. Such studies are now of vivid interest in monolayers of semiconducting transition metal dichalcogenides (S-TMDs)\cite{Cao_NatCommun_2012, Zeng_NatNano_2012, Mak_NatNano_2012, Jones_NatNano_2013, Qinsheng_ACSNano_2013, Cong_NanoLett_2014, Wang_PRB_2014, Zhu_PRB_2014}, in which the circular polarization of light ($\sigma^\pm$) is coupled to the valley degree of freedom ($\pm K$). The expected robustness of the valley pseudospin\cite{Xiao_PRL_2012, Xu_NatPhys_2014} (due to strong spin-orbit interaction in S-TMDs), together with the possibility of its optical orientation at room temperature\cite{Sallen_PRB_2012,Sanfeng_ACSNano_2013,Lagarde_PRL_2014} are promising for designing opto-valleytronic devices. Unfortunately, there exists an efficient mechanism of disorientation of the valley pseudospin. It is due to strong electron-hole exchange interaction\cite{Glazov_PRB_2014, Yu_PRB_2014, Dery_PRB_2015, Yan_SciRep_2015, Glazov_PSSB_2015}, which is particularly effective when the excited electron-hole pairs occupy the bright excitonic states. The character of the ground exciton states in S-TMDs monolayers can be, however, very different, depending on the actual order of spin-orbit split bands in the conduction band\cite{Kromanyos_2DMaterials_2015, Dery_PRB_2015}. Whereas the ground state of the exciton appears to be optically active in monolayers of molybdenum dichalcogenides, in tungsten dichalcogenides it is, to the first approximation, electric-dipole forbidden (optically dark). The latter material systems may therefore be expected to display a significantly suppressed relaxation of the valley pseudospin.

In search for inefficient relaxation of the valley pseudospin we have studied monolayers of tungsten diselenide (WSe$_2$) which were mechanically exfoliated from bulk crystals and transferred on Si/SiO$_2$ substrates (see Methods). Optical orientation experiments have been carried out in continuous-wave (CW) and time-resolved operational modes, at low temperatures (${T\sim5}$~K), and as a function of the magnetic field~(see Methods). A set of our representative results from the CW experiments is presented in Fig.~1. A familiar low-temperature photoluminescence (PL) spectrum of the WSe$_2$ monolayer can be found in Fig.~1a. It exhibits a well-established emission pattern\cite{Jones_NatNano_2013, Wang_PRB_2014, Zhu_PRB_2014, Aivazian_NP_2015, Srivastava_NP_2015} consisting of several relatively broad ($\sim$~20~meV linewidth) peaks related to the recombination of different excitons. The peak appearing at the highest energy (1750~meV) is attributed to the optically active excitonic resonance (X). This resonance is extremely short-lived (see Supplementary Information -- SI) and becomes visible in the spectrum as a hot luminescence since the bright exciton state in a WSe$_2$ monolayer is energetically higher than the dark exciton ground state. A fast dynamics is also characteristic of the X$^-$~feature (see SI), which we attribute to the recombination of the negatively charged complex associated to the excited bright exciton state. In the CW PL spectrum, the X$^-$~peak appears as a small shoulder, which is due to a partial overlap with a stronger low-energy emission band consisting of a few peaks tentatively assigned in the literature to ``localized excitons'' (LEs)\cite{Wang_PRB_2014, Zhu_PRB_2014, Mitioglu_NanoLetter_2015, Koperski_NatNano_2015, Srivastava_NatNano_2015, Arora_Nanoscale_2015}.
\begin{figure*}
\includegraphics{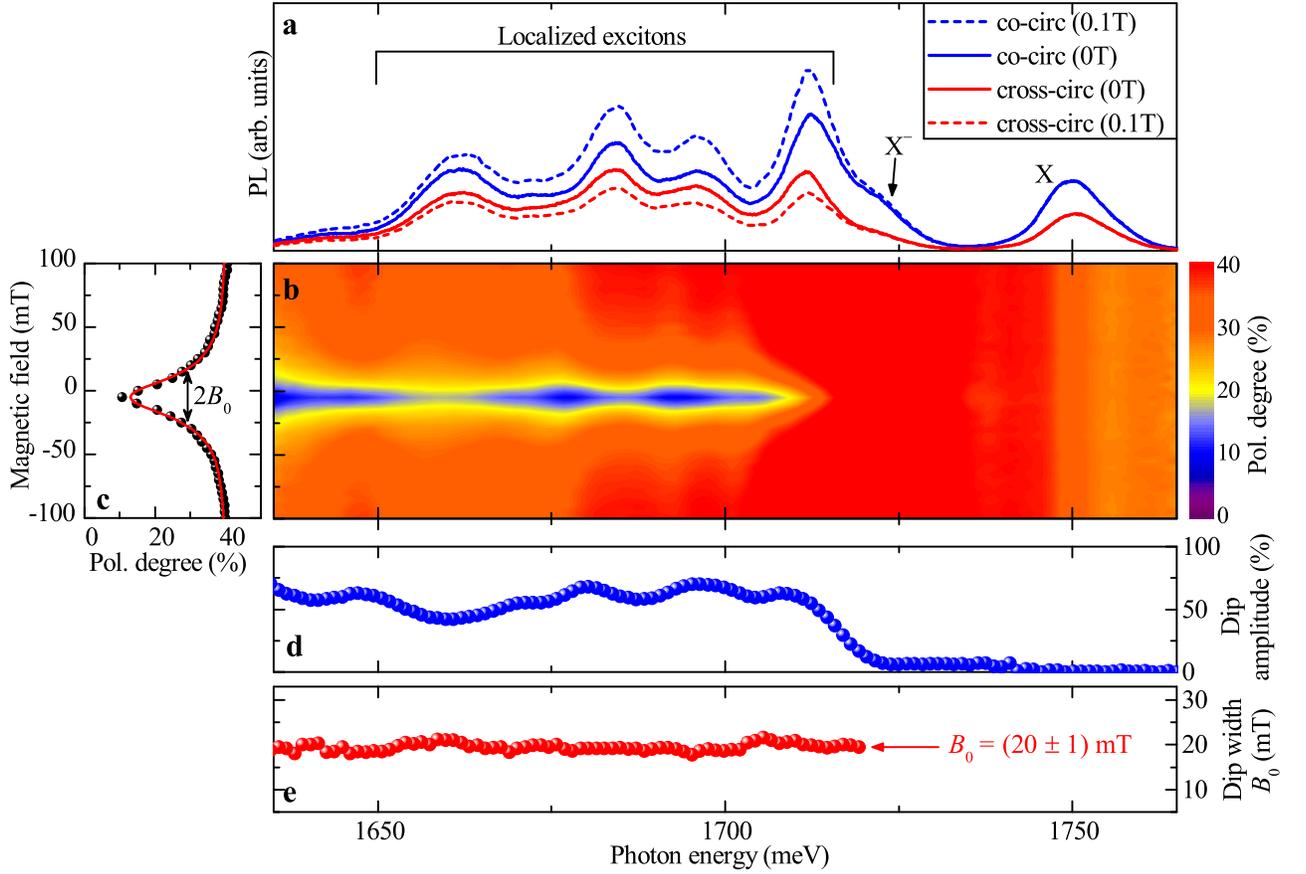}
\caption{\textbf{Tuning the valley polarization of localized excitons with a tiny magnetic field.} \textbf{a} PL spectra of a WSe$_2$ monolayer (at $T=6.5$~K) excited with the $\sigma^-$ polarized light at $E=1915$~meV. The spectra were measured at zero-field (solid lines) and at magnetic field of 100~mT applied in the Faraday geometry (dashed lines). For each field, the spectra were detected in the two circular polarizations of opposite helicities (as indicated). \textbf{b} Color-scale map presenting the circular polarization degree of monolayer WSe$_2$ PL as a function of magnetic field and emission energy. The plotted degree is obtained as an average of the two polarization degrees determined under the $\sigma^-$ and $\sigma^+$ polarized excitation. A pronounced decrease of the polarization degree at zero field is visible in the energy range corresponding to the emission of localized excitons. \textbf{c} Cross-section presenting the magnetic field dependence of the circular polarization degree measured at the energy $E=1695$~meV corresponding to one of the localized excitons. The solid line represents a fit with a formula $\mathcal{P}(B)\propto1-\alpha/\left[1+(B-B_\mathrm{rem})^2/B_0^2\right]$ accounting for the lorentzian-like field dependence of the valley pseudospin relaxation rate in line with similar effects observed for real spins in conventional semiconductors\cite{Dyakonov_JETP_1973,Berkovits_JETP_1974,optical_orientation_1984, Maialle_PRB_1993}. The parameter $\alpha$ corresponds to the percentage amplitude of the magnetic-field-induced dip in the polarization degree, $B_0$ represents the HWHM of the dip, while $B_\mathrm{rem}$ accounts for the remanent field of a superconducting coil. \textbf{d,e} The amplitude $\alpha$ (\textbf{d}) and HWHM $B_0$ (\textbf{e}) of the dip in the polarization degree determined for different emission energies.}
\end{figure*}
These peaks decay rather slowly (see SI), and we speculate that they indeed may arise from some sort of trapping effects, though involving the dark excitons. These latter excitons hardly recombine alone, but presumably can annihilate with the emission of a photon, for example, after forming a more complex excitonic object\cite{Dery_PRB_2015}.

Following the approach introduced in earlier studies\cite{Cao_NatCommun_2012, Zeng_NatNano_2012, Mak_NatNano_2012, Jones_NatNano_2013, Qinsheng_ACSNano_2013, Cong_NanoLett_2014, Wang_PRB_2014, Zhu_PRB_2014}, we investigate the pseudospin properties using polarization-resolved PL measurements. The relevant quantity in such experiments is the circular polarization degree of the emitted light defined as $\mathcal{P}=(I_\mathrm{co}-I_\mathrm{cross})/(I_\mathrm{co}+I_\mathrm{cross})$, where $I_\mathrm{co}$ ($I_\mathrm{cross}$) is the PL intensity detected in the circular polarization of the same (opposite) helicity as the excitation light. Such a quantity is of principal interest for valleytronics, since it directly reflects the degree of pseudospin conservation in S-TMDs monolayers. Fig.~1b shows the dependence of the PL polarization degree on the magnetic field applied in the Faraday geometry. The data clearly demonstrate the effect of the enhancement of the PL polarization upon application of the magnetic field, which appears in the energy range corresponding to the LE emission band. After switching on either positive or negative field, the polarization degree of the LEs increases and saturates at a significantly higher level, thus showing a zero-field dip when plotted against the field value (see Fig.~1c). Remarkably, the observed change in the polarization degree is qualitatively distinct from the recently reported changes of the PL polarization at high magnetic fields, mainly caused by different populations of the Zeeman-split excitonic levels\cite{Aivazian_NP_2015, Wang_2DMaterials_2015, Mitioglu_NanoLetter_2015}. In contrast, the effect reported here is due to the influence of small external magnetic field on the efficiency of one of the relevant processes of the pseudospin relaxation in our system. Following a large body of optical orientation studies in conventional semiconductors\cite{optical_orientation_1984, spin_physics_2008} we consider that this particular process of the pseudospin depolarization can be understood in terms of the presence of fictitious in-plane magnetic fields\cite{Dyakonov_JETP_1971, Maialle_PRB_1993}, such as, for example, those resulting from the electron-hole exchange interaction in S-TMDs monolayers \cite{Glazov_PRB_2014, Yu_PRB_2014, Glazov_PSSB_2015}. The effective depolarization rate ($\gamma_\mathrm{depol}$) related to such fictitious fields can be suppressed by the application of the external magnetic field directed perpendicularly to the monolayer plane\cite{Dyakonov_JETP_1973,Berkovits_JETP_1974}. The magnitude $B$ of the field needed for such a suppression is given by the condition $g \mu_B B/\hbar \gamma_\mathrm{depol}\approx 1$. Setting $B \approx 20$~mT as observed in the experiment (see Figs.~1b,c) we estimate $1/\gamma_\mathrm{depol}$ to be between $35$~ps and $140$~ps using g-factor values between $g=4$ and $g=12$ that have been reported in the literature for different PL transitions in WSe$_2$ monolayers\cite{Aivazian_NP_2015, Srivastava_NP_2015,Li_PRL_2014, MacNeil_PRL_2015, Wang_2DMaterials_2015,Koperski_NatNano_2015}. In any case, the estimated depolarization time significantly exceeds the pseudospin relaxation times of optically active excitons in S-TMDs monolayers that have been reported until now.

As seen in Fig.~1, the field-induced enhancement of the polarization degree occurs only for the localized excitons. The relative amplitude of this enhancement (or, equivalently, the amplitude of the polarization dip at $B=0$) is slightly different for each LE (Fig.~1d), however it always exceeds 40\%. On the other hand, in spite of different optical responses of each LE peak at high magnetic fields (e.g., different $g$-factors\cite{Koperski_NatNano_2015}), the width (HWHM) $B_0$ of the polarization dip is practically independent of the emission energy and yields about 20~mT for all LEs (Fig.~1e). This finding is of special importance, since it directly implies that the field-modulated depolarization of the valley pseudospin occurs at the same level for all transitions. It cannot happen at the energy level at which the carriers are optically injected, since we observe the same effect to appear independently of the excitation power or energy, including the laser excitations set either above or below the $B$-exciton resonance (see SI for details). Therefore, the effect must occur at an intermediate state, which we assume to be the dark ground state of the neutral exciton. Such an attribution also explains the low efficiency of the pseudospin relaxation, since dark excitons in S-TMDs monolayers should be rather weakly affected by depolarization effects arising due to the electron-hole exchange interaction\cite{Yu_PRB_2014, Glazov_PSSB_2015}.

Our interpretation is further supported by the analysis of the orientation of the magnetic field needed to suppress the relaxation. The experimental results from Fig.~2a clearly demonstrate that the dip in the polarization degree becomes wider when the magnetic field is applied along a direction tilted by the angle $\alpha_B$ from the out-of-WSe$_2$-plane direction. A clear correlation between the inverse HWHM of the polarization dip $1/B_0$ and $\cos(\alpha_B)$ inferred from our experiments (Fig.~2c) unequivocally proves that the observed field-modulation of the pseudospin relaxation occurs only due to the out-of-plane component of the magnetic field. Such magnetic anisotropy is characteristic of the valley Zeeman effect.

\begin{figure}
\includegraphics{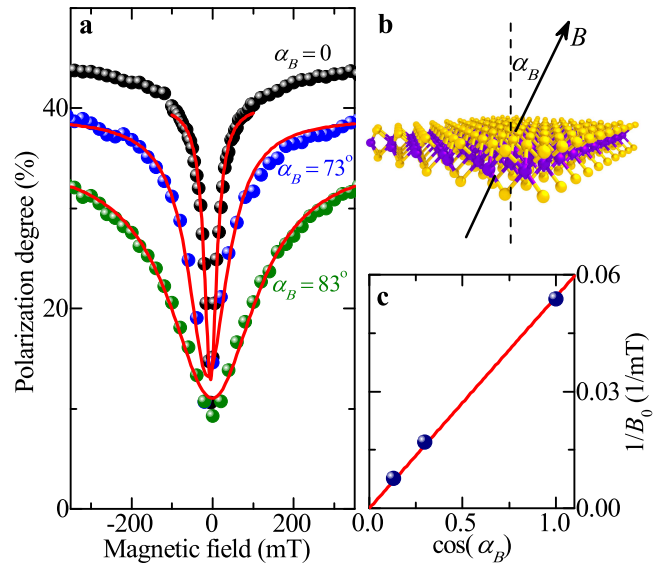}
\caption{\textbf{Magnetic anisotropy of the field-induced decrease of the valley polarization.} \textbf{a} The circular polarization degree of the WSe$_2$ monolayer PL measured vs the magnetic field at the emission energy $E=1695$~meV related to one of the localized excitons. Each set of data points corresponds to a different orientation of the magnetic field, which is quantified by the angle $\alpha_B$ between the field vector and an the out-of-plane direction (as schematically depicted in \textbf{b}). The angles $\alpha_B$ were determined from the comparison between the magnitudes of high-field Zeeman splittings of excitonic lines measured at tilted magnetic field and at the field applied perpendicularly to the surface of the sample. \textbf{c} The inverse width $1/B_0$ of the magnetic-field-induced dip in the circular polarization degree as a function of $\cos(\alpha_B)$. The solid line represents the linear fit.}
\end{figure}

\begin{figure*}
\includegraphics{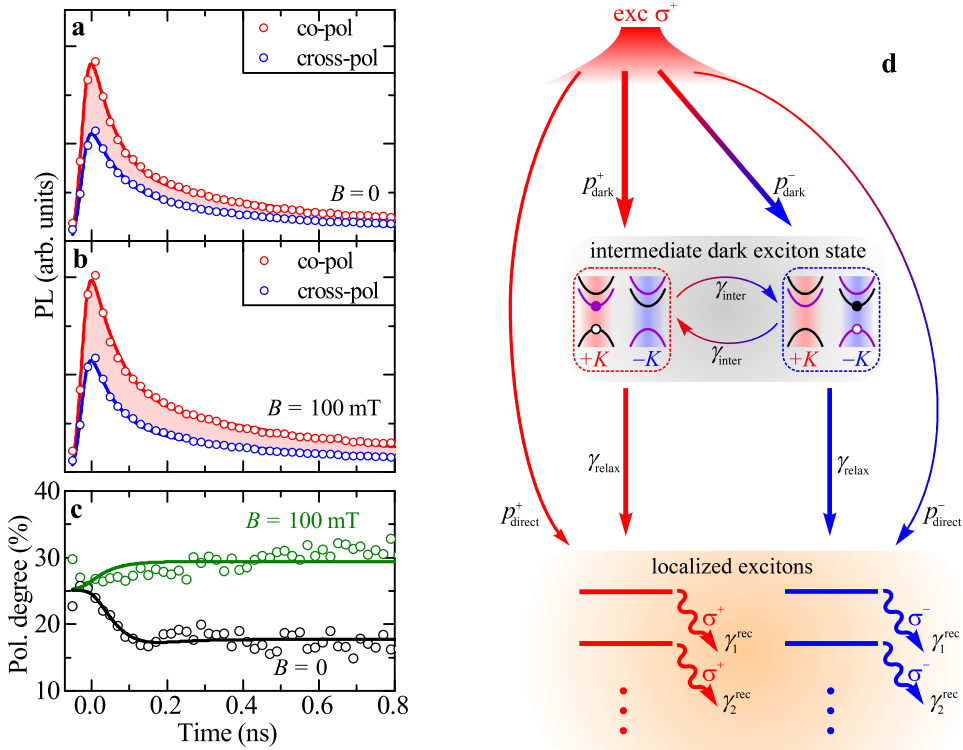}
\caption{\textbf{Valley polarization enhancement due to quenching of the inter-valley scattering by a tiny magnetic field.} \textbf{a,b} Time-resolved PL of the localized excitons integrated over the emission energy ranging from $E=1650$~meV to $E=1710$~meV. The temporal profiles were measured under the $\sigma^+$ polarized pulsed excitation at $E=1910$~meV in two circular polarizations of detection and for two values of the magnetic field: $B=0$ (\textbf{a}) and $B=100$~mT (\textbf{b}). \textbf{c} The circular polarization degree of the localized excitons emission as a function of time after the laser pulse for $B=0$ and $B=100$~mT. The solid lines in \textbf{a}, \textbf{b} and \textbf{c} represent the fit to the experimental data with the rate-equation model described in the text. \textbf{d} A scheme of the states and transitions included in the rate-equation model (a quantitative description of the model, together with the list of used parameters is provided in SI). The horizontal lines indicate the states, while the arrows mark the transitions. The color indicates the valley polarization (red -- $\sigma^+$, blue -- $\sigma^-$).}
\end{figure*}

More information on the pseudospin dynamics in our system is provided by time-resolved photoluminescence measurements performed under pulsed $\sigma^+$ polarized excitation. Figs.~3a,b show representative circular-polarization-resolved PL decay profiles measured at zero magnetic field and a tiny field of 100~mT. These profiles represent the temporal evolution of the PL signal integrated over the entire energy range corresponding to the LEs emission. All of the presented time traces exhibit a similar shape featuring a relatively rapid rise followed by a slower, multi-exponential decay. Nonetheless, the polarization degrees inferred from the pairs of profiles measured at different fields demonstrate drastically different temporal evolution, as seen in Fig.~3c. In the absence of the magnetic field the important loss of the polarization degree is clearly visible on the time scale of $\sim$~100~ps after the laser pulse, and afterwards ($> $~200~ps) $\mathcal{P}$ sets at practically constant level. Remarkably, the initial loss of the polarization degree is fully suppressed upon the application of a tiny magnetic field ($B=100$~mT), when $\mathcal{P}$ remains constant in time but at the significantly higher level.

The scheme of the minimal rate-equation model which accounts for our time-resolved data is presented in Fig.~3d. We assume that the electron-hole pairs initially photo-created by a $\sigma^+$ polarized laser pulse in the $+K$ valley relax instantaneously, in part towards the intermediate dark exciton state and in another part directly towards the LEs states. The initial populations of the LE and dark excitonic states are denoted by $p_\mathrm{direct}^\pm$ and $p_\mathrm{dark}^\pm$, respectively, where $\pm$ accounts for different $\pm K$ valley occupations. Particularly important are the populations of dark excitonic states, which are crucial for the modelled polarization effects. The initial valley polarization of dark excitons ($p_\mathrm{dark}^+ > p_\mathrm{dark}^-$ under $\sigma^+$ excitation) decays with the rate of $\gamma_\mathrm{inter}$ due to the inter-valley scattering processes. In parallel, dark excitons relax further towards the localized excitons (and/or form more complex excitonic objects) with the rate of $\gamma_\mathrm{relax}$. We assume that the localized excitons are no longer subjected to efficient inter-valley scattering, in accordance with previous studies of S-TMDs monolayers\cite{Wang_PRB_2014,Crooker_NatCom_2015} and in line with negligible decay of the polarization degree on the timescale of $0.5-1$~ns in our experiment. For the sake of simplicity, in our model we take into account only two types of the localized excitons with different lifetimes. This is a minimal assumption required to account for the non-exponential character of the temporal PL profiles shown in Figs.~3a,b. This simplification does not substantially influence the overall character of the calculated transients of the polarization degree due to a suppression of the inter-valley scattering at the level of localized excitons. The efficiency $\gamma_\mathrm{inter}$ of the inter-valley scattering of dark excitons is the only parameter assumed to be affected by the magnetic field. Using $1/\gamma_\mathrm{inter}=120$~ps in the absence of the magnetic field and fixing $\gamma_\mathrm{inter}=0$ at $B=100$~mT we have well reproduced our time-resolved data (see solid lines in Figs.~3a,b,c and SI for more details on the data simulation procedures). The extracted value of $1/\gamma_\mathrm{inter}=1/120$~ps is fully consistent with our preliminary conclusions and the estimation of $1/\gamma_\mathrm{depol}\approx35 \div 140$~ps for the field-tunable rate of the pseudospin relaxation in a WSe$_2$ monolayer deduced from the CW PL experiments.

Concluding, our continuous-wave and time-resolved optical orientation studies uncover a novel channel for the relaxation of the exciton pseudospin in a WSe$_2$ monolayer. The efficiency of this process is about two orders of magnitude weaker than that previously reported for optically active excitons in S-TMDs monolayers. Remarkably, this relaxation process can be completely switched off by the application of a tiny magnetic field of the order of $100$~mT. The latter finding invokes an interesting possibility to modulate (at the level of up to $50\%$) the polarization degree of the emitted light using, for example, the electric micro-loop wrapped around the WSe$_2$ monolayer flake. A phenomenological model invoking the role of the dark excitonic state in optical properties of the WSe$_2$ monolayer is presented to account for the present experimental finding. We believe that these findings will also challenge further investigations towards better understanding and possible practical explorations of the opto-valleytronics concepts in thin layers of S-TMDs.

\vspace{\baselineskip}
\noindent
\textbf{Methods}\\
The samples studied in our experiments were obtained by mechanical exfoliation of bulk WSe$_2$. In order to fabricate monolayer flakes of high quality and sufficient size for optical probing, we used a two-step approach in which the flakes were first filed down with the help of chemically pure backgrinding tape and then transferred onto a Si/(90~nm) SiO$_2$ substrate by means of polydimethylsiloxane-based DGL-X8 elastomeric films from Gel-Pak. To improve the cleanliness and stickiness of the target substrate we ashed it with oxygen plasma shortly before the final stage of the exfoliation process. The atomic-force-microscopy (AFM) characterization of the samples was performed with the aid of an NSV-VEECO-D3100 microscope operated in tapping mode under ambient conditions.

The polarization-resolved measurements were carried out in a high-resolution $\mu$PL setup on a sample placed in a helium gas ($T=6.5$~K) inside a magneto-optical cryostat. The cryostat was equipped with a split-coil superconducting magnet permitting for optical experiments in either the Faraday or Voigt geometry. A high numerical aperture of the cryostat allowed to perform the optical experiments even when the cryostat was intentionally rotated about its vertical axis by up to 20$^\circ$, which we utilized to apply the magnetic field along an oblique direction with respect to the sample surface. The PL was excited either by a continuous-wave diode lasers (at 488~nm or 647~nm) or by femtosecond pulses from a frequency-doubled tunable optical-parametric-oscillator (OPO) synchronously pumped by a mode-locked Ti:sapphire laser. The laser beam was focalized to a spot of diameter smaller than 2~$\mu$m by an aspheric lens immersed in the helium gas together with the sample. The lens was mounted on piezo-electric $x-y-z$ stages allowing to scan over the sample surface with a submicrometre precision. The PL signal from the sample was collected by the same lens and dispersed with the monochromator. The PL spectra were recorded by a charge-coupled device (CCD) camera or, in case of time-resolved measurements, by a synchroscan Hamamatsu streak camera. The polarization control of both the exciting and detected light was realized with a set of polarization optics (including a linear polarizer, $\lambda/2$ and $\lambda/4$ waveplates) placed either in the laser or the signal beam.

\vspace{\baselineskip}
\noindent
\textbf{Acknowledgments}\\
The authors thank J. Marcus for his contributions in the initial stage of this work, are grateful to L. Vina for valuable discussions, and acknowledge the support from the European Research Council (MOMB project no. 320590), the EC Graphene Flagship project (no. 604391), and the Polish National Science Center under decision DEC-2013/10/M/ST3/00791. One of us (T.S.) was supported by the Foundation for Polish Science (FNP).\\

\noindent \textbf{Author contributions}\\
T.S. with M.G., M.K., C.F., and T.K. performed magneto-optical and time-resolved experiments, assisted by P.K. ~~T.S. analyzed the data under supervision of P.K. and M.P. ~~M.P. proposed the phenomenological interpretation of the experiment, while T.S. carried out the modelling of the data assisted by T.K.~~ K.N. fabricated the exfoliated samples and performed their AFM characterization. P.K. and M.P. supervised the project. T.S., T.K. and M.P. prepared the manuscript in consultation with all other authors.

\vspace{\baselineskip}
\noindent
\textbf{Supplementary Information} accompanies this paper.

\vspace{\baselineskip}
\noindent
\textbf{Competing financial interests:} The authors declare no competing financial interests.

\end{document}


\title{Supplementary Information for ``Tuning valley polarization in a WSe$_2$ monolayer with a tiny magnetic field''}

\author{T.~\surname{Smole\'nski}$^{1}$}
\author{M.~\surname{Goryca}$^{1}$}
\author{M.~\surname{Koperski}$^{1,2}$}
\author{C.~\surname{Faugeras}$^{2}$}
\author{T.~\surname{Kazimierczuk}$^{1}$}
\author{K.~\surname{Nogajewski}$^{2}$}
\author{P.~\surname{Kossacki}$^{1}$}\email{Piotr.Kossacki@fuw.edu.pl}
\author{M.~\surname{Potemski}$^{2}$,}\email{Marek.Potemski@lncmi.cnrs.fr}

\affiliation{
$^{1}$ Institute of Experimental Physics, Faculty of Physics, University of Warsaw, ul. Pasteura 5, 02-093 Warsaw, Poland\\
$^{2}$ Laboratoire National des Champs Magn\'etiques Intenses, CNRS-UGA-UPS-INSA-EMFL, 25 Rue des Martyrs, Grenoble 38042, France}
\maketitle

\begin{center}
\textbf{Identification of emission peaks visible in a PL spectrum of monolayer WSe$_2$}
\end{center}

For all studied monolayer WSe$_2$ flakes the energies of free and localized excitons are partially overlapping, which hinders the identification of the emission peaks visible in a time-integrated PL spectrum. However, the peaks may be unequivocally identified in a time-resolved experiment by taking advantage of a drastically different PL decay dynamics of both types of excitons. Fig. S1a shows temporally gated PL spectra measured with a synchroscan streak camera in two different time windows following the excitation pulse. The spectrum just after the pulse features only two higher-energy peaks. These peaks disappear in the spectrum measured after a longer delay, which consists of several lower-energy peaks. On this basis, the two higher-energy peaks are identified as related to free excitons, while the lower-energy peaks are attributed to localized excitons exhibiting much longer decay dynamics\cite{Wang_PRB_2014, Heinz_NP_2015} (which is also seen in the PL decay profiles presented in Fig. S1b). Additionally, the comparison between the energies of the free excitons with the results of previous optical studies of monolayer WSe$_2$ \cite{Jones_NN_2012, Srivastava_NP_2015, Aivazian_NP_2015, Wang_2D_2015} allows us to identify these excitons as neutral (X) and negatively charged (X$^-$) (as marked in Fig. S1a). Note that an energy overlap between the peaks related to X$^-$ and to the highest-energy localized exciton is the underlying reason for which the X$^-$ emission appears in the time-integrated PL spectrum as a shoulder at the high-energy side of the localized excitons emission band.

\begin{figure*}[h]
\includegraphics{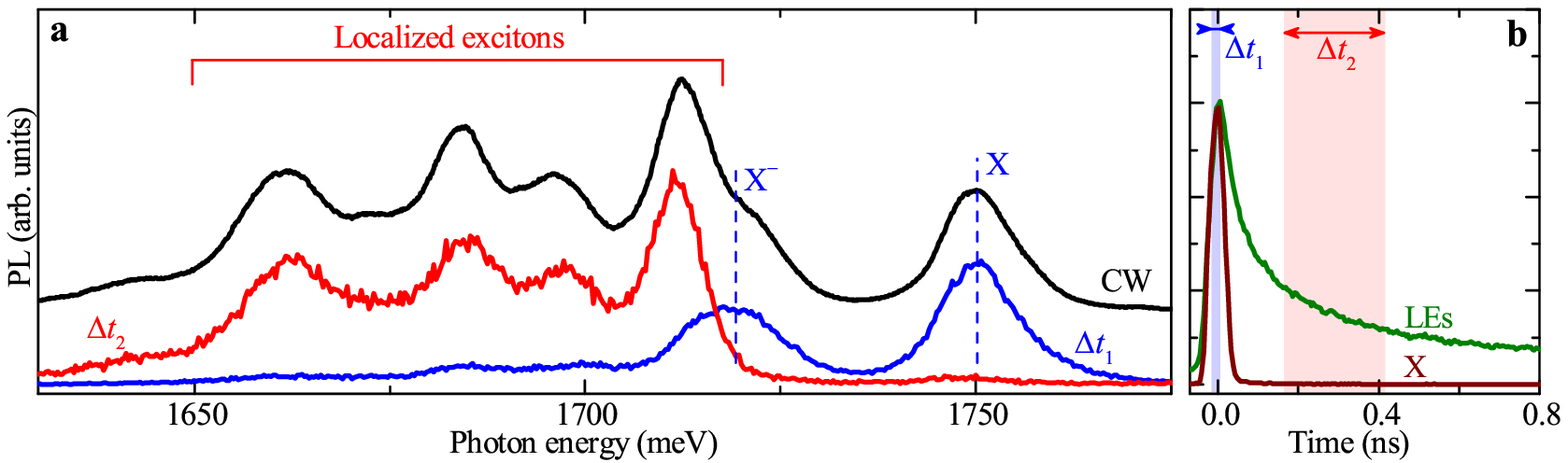}
\caption{\textbf{a} Zero-field time-gated PL spectra ($T=6.5$~K) of a single monolayer WSe$_2$ measured in two different time windows following the laser pulse: during $\Delta t_1=20$~ps at the point of laser excitation (blue line) and during $\Delta t_2=250$~ps at time 300~ps after the pulse (red line). The time-integrated PL spectrum measured under continuous-wave (CW) excitation is shown for the reference (black line). \textbf{b} The temporal decay profiles measured at two different energy ranges corresponding to the emission of free exciton X and localized excitons. Color rectangles schematically mark the time windows at which the time-gated PL spectra from \textbf{a} were obtained. For the sake of the presentation, the spectra and decay profiles from \textbf{a} and \textbf{b} were arbitrarily rescaled.}
\end{figure*}

\begin{center}
\textbf{Field-induced valley polarization enhancement for different power and energy of the optical excitation}
\end{center}

As proven by our experiments, the effect of the valley polarization enhancement by a tiny magnetic field occurs in a wide range of energies and intensities of the optical excitation. Figs. S2a and S2b present examples of the magnetic-field dependencies of the circular polarization degree measured for one of the localized excitons at two excitation powers different by almost an order of magnitude and two excitation energies lying below and above the energy $E=2.2$~eV of the $B$-exciton in monolayer WSe$_2$ \cite{Wang_PRB_2014, Arora_NanoScale_2015}. The data clearly show that neither the shape nor the width (HWHM) $B_0$ of the dip in polarization degree depend on the excitation conditions, which confirms the robustness of the observed effect. Simultaneously, the variation of the optical excitation parameters slightly modifies the relative amplitude of the dip as well as the saturation level of the polarization degree in the presence of the magnetic field, as indicated by different scales on left and right axes in Figs. S2a,b. The latter effect is particularly pronounced in Fig. S2b, which evidences much lower circular polarization degree of the localized excitons in case of high-energy excitation. This effect is most likely related to a decrease in overall efficiency of the valley pseudospin optical pumping, while the mechanism behind the polarization dip remains not affected by the excitation conditions.

\begin{figure*}
\includegraphics{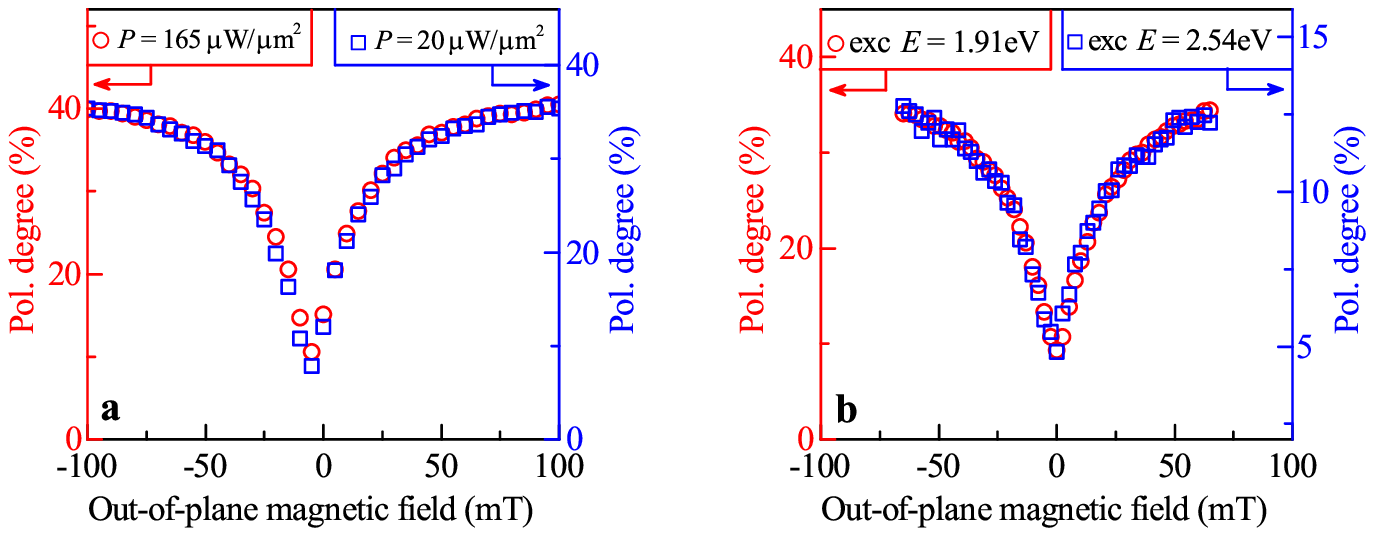}
\caption{\textbf{a,b} The circular polarization degree of monolayer WSe$_2$ PL measured as a function of out-of-plane component of the magnetic field at emission energy $E=1695$~meV related to one of localized excitons. The data were obtained for two different powers of optical excitation (\textbf{a}) and two different excitation energies (\textbf{b}). Note that in each plot the scales on the left and right vertical axes are different showing the polarization degrees of the two series of the presented data points (as indicated).}
\end{figure*}

\vspace{0.7cm}
\begin{center}
\textbf{Description of rate-equation modelling of the time-resolved PL experiments}
\end{center}

Here we provide a quantitative description of the rate-equation model used to account for the temporal PL profiles of the localized excitons measured in our experiment. In the model we include two groups of states: a single dark state and two states of localized excitons. The time-dependent populations of these states are denoted as $N_\mathrm{dark}^\pm(t)$ and $N_{\mathrm{LE},i}^\pm(t)$, respectively, where $i=1,2$ enumerates the two considered LE states, whereas the upper $\pm$ index refers to $\pm K$ valley. We assume that before the laser pulse (i.e., for $t<0$) each of these populations equals 0, while instantly after the $\sigma^+$ polarized pulse the populations of dark and localized excitons in each valley are given by
\begin{eqnarray}
N_{\mathrm{dark}}^\pm(t=0)&=&p_{\mathrm{dark}}^\pm,\\
N_{\mathrm{LE},i}^\pm(t=0)&=&\alpha_ip_{\mathrm{direct}}^\pm,\ \mathrm{for}\ i=1,2,
\end{eqnarray}
where $\alpha_i$ denotes the relative probabilities of the occupation of both considered localized excitons at $t=0$ (with the condition that $\alpha_1+\alpha_2=1$). The temporal evolution of these populations for $t>0$ is described by the following equations, which can be directly inferred from the model scheme in Fig. 3d:
\begin{eqnarray}
\frac{\mathrm{d}N_{\mathrm{dark}}^\pm(t)}{\mathrm{d}t}&=&-\gamma_\mathrm{relax}N_{\mathrm{dark}}^\pm(t)-\gamma_\mathrm{inter}N_{\mathrm{dark}}^\pm(t)+\gamma_\mathrm{inter}N_{\mathrm{dark}}^\mp(t),\\
\frac{\mathrm{d}N_{\mathrm{LE},i}^\pm(t)}{\mathrm{d}t}&=&-\gamma^\mathrm{rec}_iN_{\mathrm{LE},i}^\pm(t)+\beta_i\gamma_\mathrm{relax}N_{\mathrm{dark}}^\pm(t),\ \mathrm{for}\ i=1,2,
\end{eqnarray}
where $\gamma^\mathrm{rec}_i$ is the radiative recombination rate of $i$th localized exciton, $\gamma_\mathrm{inter}$ is the rate of inter-valley scattering between the dark excitons, $\gamma_\mathrm{relax}$ is the rate of the energy relaxation of the dark excitons (localization rate), while $\beta_i$ describes the relative probability that the relaxation of a dark exciton will result in a formation of $i$th localized exciton (with the condition that $\beta_1+\beta_2=1$). Within such a model, the overall time-dependent PL intensity of both localized excitons in $\sigma^\pm$ polarization can be expressed (up to the normalization constant) as
\begin{equation}
\mathrm{PL}_\mathrm{LE}^\pm(t)=\theta(t)\sum_{i=1,2}\gamma^\mathrm{rec}_iN_{\mathrm{LE},i}^\pm(t),
\end{equation}
where $\theta(t)$ is the Heaviside function, which equals 0 for $t<0$ and 1 for $t\ge0$. For the sake of the comparison with the experimental results, this formula is convoluted with the Gaussian profile of the width corresponding to an overall temporal resolution of our experimental setup. Finally, the formula is used to fit to the temporal profiles of the PL and polarization degree measured at $B=0$ and $B=100$~mT. All of the model parameters are taken to be independent of the magnetic field except for the rate of the inter-valley scattering $\gamma_\mathrm{inter}$, which is set to zero at $B=100$~mT. The values of all the other model parameters (including the inter-valley scattering rate $\gamma_\mathrm{inter}$ at zero magnetic field) are listed in Table I. 

\setlength\extrarowheight{5pt}
\begin{table}[h]
\begin{center}
\begin{tabular}{|>{\hspace{0.27pc}}c<{\hspace{0.27pc}}||>{\hspace{0.27pc}}c<{\hspace{0.27pc}}|>{\hspace{0.27pc}}c<{\hspace{0.27pc}}|>{\hspace{0.27pc}}c<{\hspace{0.27pc}}|>{\hspace{0.27pc}}c<{\hspace{0.27pc}}|>{\hspace{0.27pc}}c<{\hspace{0.27pc}}|>{\hspace{0.27pc}}c<{\hspace{0.27pc}}|>{\hspace{0.27pc}}c<{\hspace{0.27pc}}|>{\hspace{0.27pc}}c<{\hspace{0.27pc}}|>{\hspace{0.27pc}}c<{\hspace{0.27pc}}|>{\hspace{0.27pc}}c<{\hspace{0.27pc}}|>{\hspace{0.27pc}}c<{\hspace{0.27pc}}|>{\hspace{0.27pc}}c<{\hspace{0.27pc}}|}
\hline
Parameter & $p_\mathrm{dark}^+$ & $p_\mathrm{dark}^-$ & $p_\mathrm{direct}^+$ & $p_\mathrm{direct}^-$ & $\alpha_1$ & $\alpha_2$ & $1/\gamma^\mathrm{rec}_1$ & $1/\gamma^\mathrm{rec}_2$ & $1/\gamma_\mathrm{inter}(B=0)$ & $1/\gamma_\mathrm{relax}$ & $\beta_1$ & $\beta_2$\\[5pt]\hline\hline
Value & 0.467 & 0.249 & 0.177 & 0.107 & 0.498 & 0.502 & 42~ps & 683~ps & 120~ps & 48~ps & 0.109 & 0.891 \\[5pt]\hline
\end{tabular}
\end{center}
\caption{The values of the rate-equation model parameters used to calculate the theoretical curves presented in Figs.~3a-c.}
\end{table}